\documentclass[twocolumn,letterpaper,aps,prd,longbibliography,superscriptaddress,showpacs,nofootinbib,floatfix]{revtex4-1}

\usepackage{graphicx}	% Include figure files
\usepackage{url}

\usepackage{xspace}	% Include xspace

\newcommand{\pt}{\mbox{$p_T$}\xspace}

\newcommand{\sqs}{\mbox{$\sqrt{s}$}\xspace}

\newcommand{\pp}{\mbox{$p$$+$$p$}\xspace}

\newcommand{\pion}{\mbox{$\pi^0$}\xspace}
\newcommand{\piz}{\mbox{$\pi^{0}$}\xspace}
\newcommand{\ptp}{\mbox{$p^\uparrow$$+$$p$}\xspace}

\newcommand{\gevc}{~{\rm GeV}/c}

\newcommand{\gev}{~{\rm GeV}}
\newcommand{\mevcc}{~{\rm MeV}/c^2}

\def\fig#1{Fig.~\ref{#1}}
\def\eq#1{Eq.~(\ref{#1})}

\begin{document}

%%%%% NOTE:  Use the above macros in text and captions, but not in
%%%%%        the title or abstract.  Those two elements must stand
%%%%%        alone and not be dependent on the macros defined here.

\title{ Transverse single-spin asymmetries of midrapidity $\pi^0$ and 
$\eta$ mesons in polarized $p$$+$$p$ collisions at $\sqrt{s}=200$ GeV}

\newcommand{\abilene}{Abilene Christian University, Abilene, Texas 79699, USA}
\newcommand{\augie}{Department of Physics, Augustana University, Sioux Falls, South Dakota 57197, USA}
\newcommand{\banaras}{Department of Physics, Banaras Hindu University, Varanasi 221005, India}
\newcommand{\barc}{Bhabha Atomic Research Centre, Bombay 400 085, India}
\newcommand{\baruch}{Baruch College, City University of New York, New York, New York, 10010 USA}
\newcommand{\bnlcoll}{Collider-Accelerator Department, Brookhaven National Laboratory, Upton, New York 11973-5000, USA}
\newcommand{\bnlphys}{Physics Department, Brookhaven National Laboratory, Upton, New York 11973-5000, USA}
\newcommand{\caucr}{University of California-Riverside, Riverside, California 92521, USA}
\newcommand{\charlesczech}{Charles University, Ovocn\'{y} trh 5, Praha 1, 116 36, Prague, Czech Republic}
\newcommand{\cns}{Center for Nuclear Study, Graduate School of Science, University of Tokyo, 7-3-1 Hongo, Bunkyo, Tokyo 113-0033, Japan}
\newcommand{\colorado}{University of Colorado, Boulder, Colorado 80309, USA}
\newcommand{\columbia}{Columbia University, New York, New York 10027 and Nevis Laboratories, Irvington, New York 10533, USA}
\newcommand{\czechtech}{Czech Technical University, Zikova 4, 166 36 Prague 6, Czech Republic}
\newcommand{\debrecen}{Debrecen University, H-4010 Debrecen, Egyetem t{\'e}r 1, Hungary}
\newcommand{\elte}{ELTE, E{\"o}tv{\"o}s Lor{\'a}nd University, H-1117 Budapest, P{\'a}zm{\'a}ny P.~s.~1/A, Hungary}
\newcommand{\eszterhazy}{Eszterh\'azy K\'aroly University, K\'aroly R\'obert Campus, H-3200 Gy\"ongy\"os, M\'atrai \'ut 36, Hungary}
\newcommand{\ewha}{Ewha Womans University, Seoul 120-750, Korea}
\newcommand{\famu}{Florida A\&M University, Tallahassee, FL 32307, USA}
\newcommand{\fsu}{Florida State University, Tallahassee, Florida 32306, USA}
\newcommand{\gsu}{Georgia State University, Atlanta, Georgia 30303, USA}
\newcommand{\hiroshima}{Hiroshima University, Kagamiyama, Higashi-Hiroshima 739-8526, Japan}
\newcommand{\howard}{Department of Physics and Astronomy, Howard University, Washington, DC 20059, USA}
\newcommand{\ihepprot}{IHEP Protvino, State Research Center of Russian Federation, Institute for High Energy Physics, Protvino, 142281, Russia}
\newcommand{\illuiuc}{University of Illinois at Urbana-Champaign, Urbana, Illinois 61801, USA}
\newcommand{\inrras}{Institute for Nuclear Research of the Russian Academy of Sciences, prospekt 60-letiya Oktyabrya 7a, Moscow 117312, Russia}
\newcommand{\instpasczech}{Institute of Physics, Academy of Sciences of the Czech Republic, Na Slovance 2, 182 21 Prague 8, Czech Republic}
\newcommand{\isu}{Iowa State University, Ames, Iowa 50011, USA}
\newcommand{\jaea}{Advanced Science Research Center, Japan Atomic Energy Agency, 2-4 Shirakata Shirane, Tokai-mura, Naka-gun, Ibaraki-ken 319-1195, Japan}
\newcommand{\jeonbuk}{Jeonbuk National University, Jeonju, 54896, Korea}
\newcommand{\kek}{KEK, High Energy Accelerator Research Organization, Tsukuba, Ibaraki 305-0801, Japan}
\newcommand{\korea}{Korea University, Seoul 02841, Korea}
\newcommand{\kurchatov}{National Research Center ``Kurchatov Institute", Moscow, 123098 Russia}
\newcommand{\kyoto}{Kyoto University, Kyoto 606-8502, Japan}
\newcommand{\lawllnl}{Lawrence Livermore National Laboratory, Livermore, California 94550, USA}
\newcommand{\losalamos}{Los Alamos National Laboratory, Los Alamos, New Mexico 87545, USA}
\newcommand{\lund}{Department of Physics, Lund University, Box 118, SE-221 00 Lund, Sweden}
\newcommand{\lyon}{IPNL, CNRS/IN2P3, Univ Lyon, Université Lyon 1, F-69622, Villeurbanne, France}
\newcommand{\maryland}{University of Maryland, College Park, Maryland 20742, USA}
\newcommand{\mass}{Department of Physics, University of Massachusetts, Amherst, Massachusetts 01003-9337, USA}
\newcommand{\michigan}{Department of Physics, University of Michigan, Ann Arbor, Michigan 48109-1040, USA}
\newcommand{\muhlenberg}{Muhlenberg College, Allentown, Pennsylvania 18104-5586, USA}
\newcommand{\nara}{Nara Women's University, Kita-uoya Nishi-machi Nara 630-8506, Japan}
\newcommand{\natmephi}{National Research Nuclear University, MEPhI, Moscow Engineering Physics Institute, Moscow, 115409, Russia}
\newcommand{\newmex}{University of New Mexico, Albuquerque, New Mexico 87131, USA}
\newcommand{\nmsu}{New Mexico State University, Las Cruces, New Mexico 88003, USA}
\newcommand{\northcg}{Physics and Astronomy Department, University of North Carolina at Greensboro, Greensboro, North Carolina 27412, USA}
\newcommand{\ohio}{Department of Physics and Astronomy, Ohio University, Athens, Ohio 45701, USA}
\newcommand{\ornl}{Oak Ridge National Laboratory, Oak Ridge, Tennessee 37831, USA}
\newcommand{\orsay}{IPN-Orsay, Univ.~Paris-Sud, CNRS/IN2P3, Universit\'e Paris-Saclay, BP1, F-91406, Orsay, France}
\newcommand{\peking}{Peking University, Beijing 100871, People's Republic of China}
\newcommand{\pnpi}{PNPI, Petersburg Nuclear Physics Institute, Gatchina, Leningrad region, 188300, Russia}
\newcommand{\pusan}{Pusan National University, Pusan 46241, Korea}
\newcommand{\riken}{RIKEN Nishina Center for Accelerator-Based Science, Wako, Saitama 351-0198, Japan}
\newcommand{\rikjrbrc}{RIKEN BNL Research Center, Brookhaven National Laboratory, Upton, New York 11973-5000, USA}
\newcommand{\rikkyo}{Physics Department, Rikkyo University, 3-34-1 Nishi-Ikebukuro, Toshima, Tokyo 171-8501, Japan}
\newcommand{\saispbstu}{Saint Petersburg State Polytechnic University, St.~Petersburg, 195251 Russia}
\newcommand{\seoulnat}{Department of Physics and Astronomy, Seoul National University, Seoul 151-742, Korea}
\newcommand{\stonybrkc}{Chemistry Department, Stony Brook University, SUNY, Stony Brook, New York 11794-3400, USA}
\newcommand{\stonycrkp}{Department of Physics and Astronomy, Stony Brook University, SUNY, Stony Brook, New York 11794-3800, USA}
\newcommand{\tenn}{University of Tennessee, Knoxville, Tennessee 37996, USA}
\newcommand{\titech}{Department of Physics, Tokyo Institute of Technology, Oh-okayama, Meguro, Tokyo 152-8551, Japan}
\newcommand{\tsukuba}{Tomonaga Center for the History of the Universe, University of Tsukuba, Tsukuba, Ibaraki 305, Japan}
\newcommand{\vandy}{Vanderbilt University, Nashville, Tennessee 37235, USA}
\newcommand{\weizmann}{Weizmann Institute, Rehovot 76100, Israel}
\newcommand{\wigner}{Institute for Particle and Nuclear Physics, Wigner Research Centre for Physics, Hungarian Academy of Sciences (Wigner RCP, RMKI) H-1525 Budapest 114, POBox 49, Budapest, Hungary}
\newcommand{\yonsei}{Yonsei University, IPAP, Seoul 120-749, Korea}
\newcommand{\zagreb}{Department of Physics, Faculty of Science, University of Zagreb, Bijeni\v{c}ka c.~32 HR-10002 Zagreb, Croatia}
\affiliation{\abilene}
\affiliation{\augie}
\affiliation{\banaras}
\affiliation{\barc}
\affiliation{\baruch}
\affiliation{\bnlcoll}
\affiliation{\bnlphys}
\affiliation{\caucr}
\affiliation{\charlesczech}
\affiliation{\cns}
\affiliation{\colorado}
\affiliation{\columbia}
\affiliation{\czechtech}
\affiliation{\debrecen}
\affiliation{\elte}
\affiliation{\eszterhazy}
\affiliation{\ewha}
\affiliation{\famu}
\affiliation{\fsu}
\affiliation{\gsu}
\affiliation{\hiroshima}
\affiliation{\howard}
\affiliation{\ihepprot}
\affiliation{\illuiuc}
\affiliation{\inrras}
\affiliation{\instpasczech}
\affiliation{\isu}
\affiliation{\jaea}
\affiliation{\jeonbuk}
\affiliation{\kek}
\affiliation{\korea}
\affiliation{\kurchatov}
\affiliation{\kyoto}
\affiliation{\lawllnl}
\affiliation{\losalamos}
\affiliation{\lund}
\affiliation{\lyon}
\affiliation{\maryland}
\affiliation{\mass}
\affiliation{\michigan}
\affiliation{\muhlenberg}
\affiliation{\nara}
\affiliation{\natmephi}
\affiliation{\newmex}
\affiliation{\nmsu}
\affiliation{\northcg}
\affiliation{\ohio}
\affiliation{\ornl}
\affiliation{\orsay}
\affiliation{\peking}
\affiliation{\pnpi}
\affiliation{\pusan}
\affiliation{\riken}
\affiliation{\rikjrbrc}
\affiliation{\rikkyo}
\affiliation{\saispbstu}
\affiliation{\seoulnat}
\affiliation{\stonybrkc}
\affiliation{\stonycrkp}
\affiliation{\tenn}
\affiliation{\titech}
\affiliation{\tsukuba}
\affiliation{\vandy}
\affiliation{\weizmann}
\affiliation{\wigner}
\affiliation{\yonsei}
\affiliation{\zagreb}
\author{U.A.~Acharya} \affiliation{\gsu} 
\author{C.~Aidala} \affiliation{\michigan} 
\author{Y.~Akiba} \email[PHENIX Spokesperson: ]{akiba@rcf.rhic.bnl.gov} \affiliation{\riken} \affiliation{\rikjrbrc} 
\author{M.~Alfred} \affiliation{\howard} 
\author{V.~Andrieux} \affiliation{\michigan} 
\author{N.~Apadula} \affiliation{\isu} 
\author{H.~Asano} \affiliation{\kyoto} \affiliation{\riken} 
\author{B.~Azmoun} \affiliation{\bnlphys} 
\author{V.~Babintsev} \affiliation{\ihepprot} 
\author{N.S.~Bandara} \affiliation{\mass} 
% ...
\collaboration{PHENIX Collaboration} \noaffiliation

\date{\today}

%------------------------------------------------------------------------------|

\begin{abstract}

%\linenumbers

We present a measurement of the transverse single-spin asymmetry for 
$\pi^0$ and $\eta$ mesons in $p^\uparrow$$+$$p$ collisions in the 
pseudorapidity range $|\eta|<0.35$ and at a center-of-mass energy of 200 
GeV with the PHENIX detector at the Relativistic Heavy Ion Collider. In 
comparison with previous measurements in this kinematic region, these 
results have factor-of-three-smaller uncertainties. As hadrons, $\pi^0$ 
and $\eta$ mesons are sensitive to both initial- and final-state 
nonperturbative effects for a mix of parton flavors. Comparisons of the 
differences in their transverse single-spin asymmetries have the 
potential to disentangle the possible effects of strangeness, isospin, 
or mass. These results can constrain the twist-3 trigluon collinear 
correlation function as well as the gluon Sivers function.

\end{abstract}

\maketitle

\section{Introduction}

Spin-momentum correlations in hadronic collisions have attracted 
increasing experimental and theoretical interest in the past two 
decades.  In particular, transverse single-spin asymmetries (TSSAs) have 
been one of the primary means to probe transverse partonic dynamics in 
the nucleon. In the context of proton-proton collisions, one 
transversely-polarized proton collides with another unpolarized proton 
and the TSSA measures the asymmetry in yields of particles that travel 
to the left versus the right of the polarized-proton-going direction. 
Large azimuthal asymmetries of up to $\approx\,$40\% have been observed 
from transversely polarized $ p^\uparrow + p $ collisions in light meson 
production at large Feynman-x ($ x_F = 2 p_L/\sqrt{s}$), from center of 
mass energies of \( \sqrt{s} = 4.9 \gev \) up to 
$500~\gev$~\cite{Klem:1976ui,Adams:1991cs,Abelev:2008af,Arsene:2008aa,Adare:2013ekj,Adam:2020edg}.  
Next-to-leading order perturbative quantum chromodynamics (QCD) 
calculations that only include spin-momentum correlations from parton 
scattering predict small asymmetries on the order of 
$m_q/Q$~\cite{Kane:1978nd}, where $m_q$ is the bare quark mass and $Q$ 
is the hard scale, indicating that significant nonperturbative effects 
must dominate the large measured asymmetries.  Two different approaches 
have been proposed to describe the large asymmetries observed in 
hadronic interactions.

%%% xyx flag predict above

In the first approach, nonperturbative parton distribution functions 
(PDFs) and fragmentation functions (FFs) are explicitly dependent on 
transverse momentum in the transverse-momentum-dependent (TMD) 
framework.  These functions depend on a soft ($k_T$) and hard ($Q$) 
momentum scale such that $\Lambda_{QCD}\lesssim k_T\ll Q$. One possible 
origin of the large TSSAs is the Sivers TMD PDF~\cite{Sivers:1989cc}, 
which correlates the nucleon transverse spin with the parton transverse 
momentum, $k_T$. Another possible origin of the TSSA is the Collins TMD 
FF~\cite{Collins:1992kk}, which correlates the transverse polarization 
of a fragmenting quark to the angular distribution of hadrons.
 
The second approach to describe the large asymmetries relies on 
collinear higher-twist effects with multiparton correlations. In the 
twist-3 approach, interference arises between scattering amplitudes 
with one and two collinear partons, which leads to a nonzero TSSA. 
This approach applies to observables in which only one sufficiently 
hard momentum scale is measured, such that $Q \gg 
\Lambda_{QCD}$~\cite{Cammarota:2020qcw}. To keep the multiparton 
correlation functions process independent, the initial- and 
final-state interactions between the struck parton and the proton remnants 
are included in the hard perturbative part of the twist-3 
collinear factorization~\cite{Aybat:2011ge}. Collinear twist-3 
correlation functions are split into two types: the quark-gluon-quark 
functions (\textit{qgq}) and the trigluon functions (\textit{ggg}).  
In the context of initial-state effects, the \textit{qgq} functions 
describe the interference from scattering off of one quark versus 
scattering off of a gluon and a quark of the same flavor, while the 
\textit{ggg} functions capture the interference between scattering off
of one gluon versus scattering off of two. The twist-3 approach is 
well suited to describe observed inclusive forward hadron asymmetries 
because the observed hadron \pt can be used as a proxy for the hard 
scale, and unlike the TMD approach, these correlation functions do not
explicitly depend on a soft-scale transverse momentum.  However, the 
twist-3 approach has been related to $k_T$ moments of TMD PDFs and TMD
FFs and has been shown to be equivalent to the TMD approach in the 
overlapping kinematic regime~\cite{Ji:2006ub}.

Because the Sivers function is PT-odd, in order to be nonzero it must 
include a soft-gluon exchange with the proton remnant, which can occur 
before and/or after the hard partonic scattering depending on the 
process~\cite{Collins:2002kn}.  Significant nonzero asymmetries due to 
the Sivers TMD PDF have been measured in semi-inclusive deep-inelastic 
lepton-nucleon scattering 
(SIDIS)~\cite{Airapetian:2004tw,Adolph:2014zba}, where the soft-gluon 
exchange can only happen in the final state.In hadronic interactions 
where at least one final-state hadron is measured, both initial- and 
final-state interactions can play competing roles in the measured 
asymmetries; here TMD-factorization breaking has been predicted due to 
soft gluon exchanges that are possible in both the initial and final 
states simultaneously~\cite{Rogers:2010dm}. Additional leading-power 
spin asymmetries have been predicted in hadronic collisions due to this 
breakdown, without which these asymmetries would be 
subleading~\cite{Rogers:2013zha}, but further work is needed to connect 
TMD-factorization breaking to experimentally measured asymmetries.  Note 
that inclusive hadron TSSA measurements in hadronic collisions appear to 
plateau at $p_T$ up to 5 \gevc~\cite{Adare:2013ekj,Abelev:2008af} and 
have been measured to be nonzero at up to 
$p_T\,\approx7$~\gevc~\cite{Adam:2020jjg}. Recent studies in the twist-3 
framework have successfully described the \pt dependence of these 
forward asymmetries by including twist-3 effects in 
hadronization~\cite{Kanazawa:2014dca}.  The twist-3 perturbative 
prediction is that the asymmetry should eventually decrease as the hard 
scale $p_T$ continues increasing~\cite{Rogers:2013zha}.

Since the inception of the collinear twist-3 and TMD factorization 
pictures, there has been theoretical evidence that they could combine 
to form a unified picture of TSSAs in hard processes.  This concept 
was recently tested with the first simultaneous global analysis of 
TSSAs in SIDIS, Drell-Yan, $e^+ e^-$ annihilation, and proton-proton 
collisions~\cite{Cammarota:2020qcw}.  This study used quark TMD PDFs 
and FFs to describe the asymmetries in processes that are sensitive to
the soft-scale momentum, i.e. SIDIS, Drell-Yan, and $e^+ e^-$ 
annihilation.  These TMD functions were also used to calculate 
collinear twist-3 \textit{qgq} correlation functions which were 
applied to inclusive forward pion asymmetry measurements from the
Relativistic Heavy Ion Collider (RHIC).  This simultaneous description 
of TSSAs across multiple collision species indicates that all TSSAs have 
a common origin that is related to multiparton correlations.

Additional questions about the origin of the TSSAs in hadronic 
interactions remain. Forward jet measurements indicate that the TSSA 
is significantly smaller than neutral pion asymmetries at similar 
$x_F$ and \sqs~\cite{Bland:2013pkt}. Nonzero kaon and antiproton 
asymmetries observed at forward rapidities show that the measured 
asymmetries cannot be due only to proton valence quark contributions 
as naively predicted in a valence-like model, where the Sivers effect 
from sea-quarks and/or gluons is ignored, and that the fragmentation 
of quarks into hadrons in which they are not valence quarks could play
a role in the observed nonzero 
asymmetries~\cite{Arsene:2008aa,BRAHMS_200GeV}. Eta meson 
measurements, sensitive to potential effects from strange quark 
contributions, isospin, and/or hadron mass show forward asymmetries 
similar in magnitude to neutral pions~\cite{Adare:2014qzo}. At 
midrapidity at RHIC, nonzero TSSAs have been measured for charged pion
pair production~\cite{Adamczyk:2015hri,Adamczyk:2017ynk}.  Even four 
decades after the initial discovery of large TSSAs in hadronic 
interactions~\cite{Klem:1976ui}, there remain many unresolved 
questions about their origin. Therefore it is crucial to continue 
extending measurements to try to better understand the nonperturbative
dynamics which are responsible for the TSSAs in hadronic collisions. 

In this paper we report a measurement of the TSSA of $\pion$ and $\eta$ 
mesons in \ptp collisions at \sqs= 200 GeV in the midrapidity region 
$|\eta|<$ 0.35. The data was taken during the 2015 RHIC run and a total 
integrated luminosity of approximately 60 pb$^{-1}$ was collected. This 
measurement extends previous measurements from RHIC to higher $p_T$ and 
reduces the statistical uncertainties by a factor of three in the 
overlapping $p_T$ region.

%\vspace{0.5cm}
\section{Analysis}

The asymmetries are measured with transversely polarized proton 
beams where the average polarization of the clockwise beam was 
$0.58\pm0.02$ and that of the counter-clockwise beam was $0.60 \pm 
0.02$~\cite{polarimetry}.  The direction of the beam polarization 
was found to be consistent with the vertical within statistical 
uncertainties.  The polarization direction of each beam 
independently changes bunch to bunch which reduces systematic 
uncertainties associated with variations in detector performance 
with time.  The relative luminosity is the ratio of the integrated 
luminosity for bunches that were polarized in opposite directions.  It
is determined by the number of times each crossing fires a minimum-bias 
(MB) trigger and is measured to better than $10^{-4}$. 
The relative luminosity values for both the beams were limited to the 
range of 0.91 to 1.09 for all beam fills used in these measurements.  
The bunch-to-bunch changes in polarization direction also allow for 
polarization-averaged measurements and, for a single-spin asymmetry 
analysis, provide two ways to measure the TSSA with the same data 
set.  This is done by sorting the particle yields for the 
polarization directions of one beam at a time, effectively 
averaging over the polarization of the other beam.  RHIC uses eight 
different spin patterns for sequential fills which are carefully 
chosen to minimize potential effects from nonzero average polarization
of either beam. The statistically independent asymmetries measured 
from the two beams are used to verify the analysis and are averaged 
together for the final result. 

The data analysis procedure is similar to our previous 
measurements~\cite{Adare:2013ekj}. Neutral pion and eta mesons are 
reconstructed via their two-photon decays by using the midrapidity 
electromagnetic calorimeter (EMCal). The EMCal is located in two 
central arms, each covering $\Delta\phi=\pi/2$ in azimuth and 
$|\eta|<0.35$ in pseudorapidity, centered at $\phi=\pi/16$ and 
$15\pi/16$.  The EMCal comprises two different types of calorimeters: 
six sectors of sampling lead-scintillator calorimeters and two sectors
of \v{C}erenkov lead-glass calorimeters~\cite{Aphecetche:2003zr}.  The 
two calorimeter systems have different granularity 
($\delta\phi\times\delta\eta=0.011\times 0.011$ in the lead 
scintillator and $0.008\times0.008$ in the lead glass) and also 
different responses to charged hadrons, which provides important 
systematic cross checks for these measurements. A tracking system 
includes a drift chamber to measure track momentum and pad chamber 
stations to measure the charged particle hit 
position~\cite{Adcox:2003zp}. The measurement of the track 
positions in front of the calorimeter is used to veto charged particles 
from the photon sample.  The beam-beam counters (BBC) are arrays of 
quartz \v{C}erenkov radiators that surround the beam pipe and are placed 
$\pm144$ cm away from the nominal collision point.  The BBC covers full 
azimuth and $3.0<|\eta|<3.9$ in pseudorapidity. They measure the 
z-vertex position; a vertex cut of $\pm30$ cm around the nominal 
collision point is used for this analysis. The MB trigger 
requires at least one charged particle to be measured in both sides of
the BBC.  This analysis is based on the data sample selected with the 
EMCal-based high-energy-photon trigger with energy threshold of 1.5 
GeV, which is taken in coincidence with the MB trigger.

Photons are identified as clusters in the EMCal and are required to pass 
a shower profile cut which suppresses clusters from hadrons.  High-\pt 
trigger photons are paired with another photon in the same event that is 
also on the same side of the detector.  A charged track veto cut 
eliminates clusters that geometrically match with a measured charged 
track, reducing background from electrons.  The contribution of EMCal 
detector noise is reduced by a minimum energy cut of 0.5 GeV and a 
time-of-flight cut of $|TOF|< 5$ ns. The timing of the cluster is 
measured by the EMCal and the time zero reference of the event is 
provided by the BBC. Each photon pair is required to pass an energy 
asymmetry cut: $\alpha=|E_1-E_2|/(E_1+E_2)<0.8$. 
The \piz yields comprise photon pairs with invariant mass in the signal 
region $\pm 25~{\rm MeV/c^2}$ from the \piz mass peak and $\eta$ 
meson yields are measured in the range $\pm70~{\rm MeV}/c^2$ around the 
$\eta$ mass peak.

The transverse single-spin asymmetries are determined with the 
``relative luminosity" formula
\begin{equation}
A_N = \frac{1}{P \, \left<\cos(\phi)\right>} \frac{{N^{\uparrow}}-\mathcal{R}{N^{\downarrow}} }{{N^{\uparrow}}+\mathcal{R}{N^{\downarrow}}},
\label{eq:relan}
\end{equation}

\noindent which compares the yield of particles for when the beam was 
polarized up versus down.  Here $P$ is the beam polarization, $N$ refers 
to the meson yield, the arrows refer to the up ($\uparrow$) or down 
($\downarrow$) directions of beam polarization, and 
$\mathcal{R}=\mathcal{L}^\uparrow/\mathcal{L}^\downarrow$ is the 
relative luminosity.  The acceptance factor, $\left<\cos(\phi)\right>$, 
accounts for the detector azimuthal coverage, where $\phi=0$ points 
$90^\circ$ from the (vertical) spin axis.  This correction is calculated 
as a function of photon pair \pt because the diphoton azimuthal 
acceptance depends heavily on the decay angle and ranges from 0.95 at 
low \pt to 0.89 at high \pt.  The asymmetry is calculated separately for 
the two detector arms and then the average weighed by the statistical 
error is taken for the final result.  As written, \eq{eq:relan} is for 
the arm to the left of the direction of travel of the beam that is being 
taken as polarized. An overall minus sign is needed for the asymmetry of 
particle yields in the arm to the right of the polarized-beam-going 
direction.

An alternative method of calculating the asymmetry is the ``square 
root'' formula
\begin{equation}
A_N = \frac{1}{P \, \left<\cos(\phi)\right>} 
\frac{\sqrt{N_L^{\uparrow}N_R^{\downarrow}}-\sqrt{N_L^{\downarrow}N_R^{\uparrow}}}
{\sqrt{N_L^{\uparrow}N_R^{\downarrow}}+\sqrt{N_L^{\downarrow}N_R^{\uparrow}}},
\label{eq:sqrtan}
\end{equation} 

\noindent which is used as a cross check.  This formula combines data 
from the two arms (left and right) and both beam polarization directions 
(up and down). The subscripts in \eq{eq:sqrtan} refer to the yields to 
the left ($L$) and right ($R$) side of the polarized-beam-going direction.

%%%%%%%%%%%%%
% Invariant Mass             %
%%%%%%%%%%%%%

%%%%%%%%%%%%%%%%%%%%%%%%%%%%%%%%%%%%%%%%%%%%%%%%%% Fig_1
\begin{figure}[htbp]
  \includegraphics[width=1.0\linewidth]{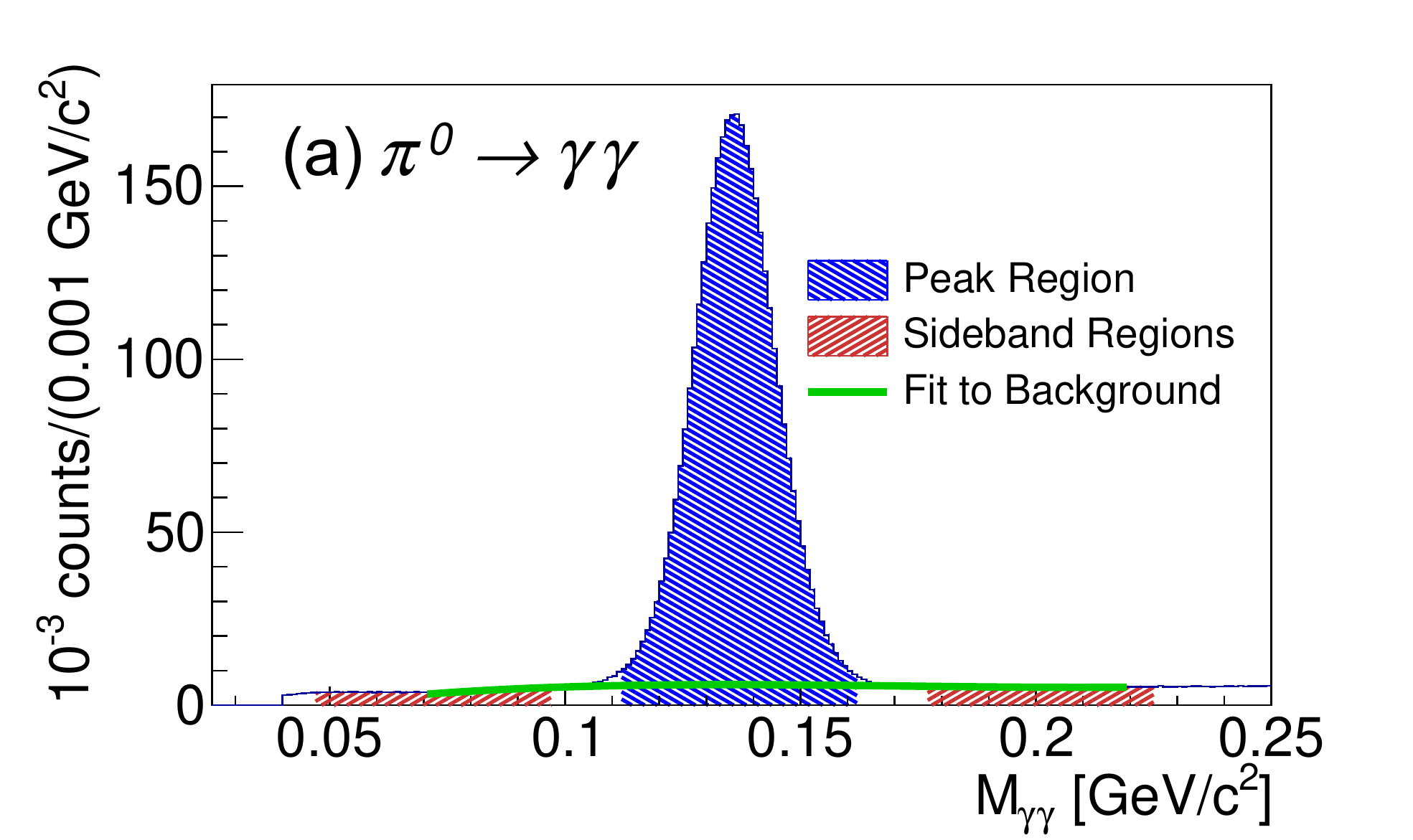}
  \includegraphics[width=1.0\linewidth]{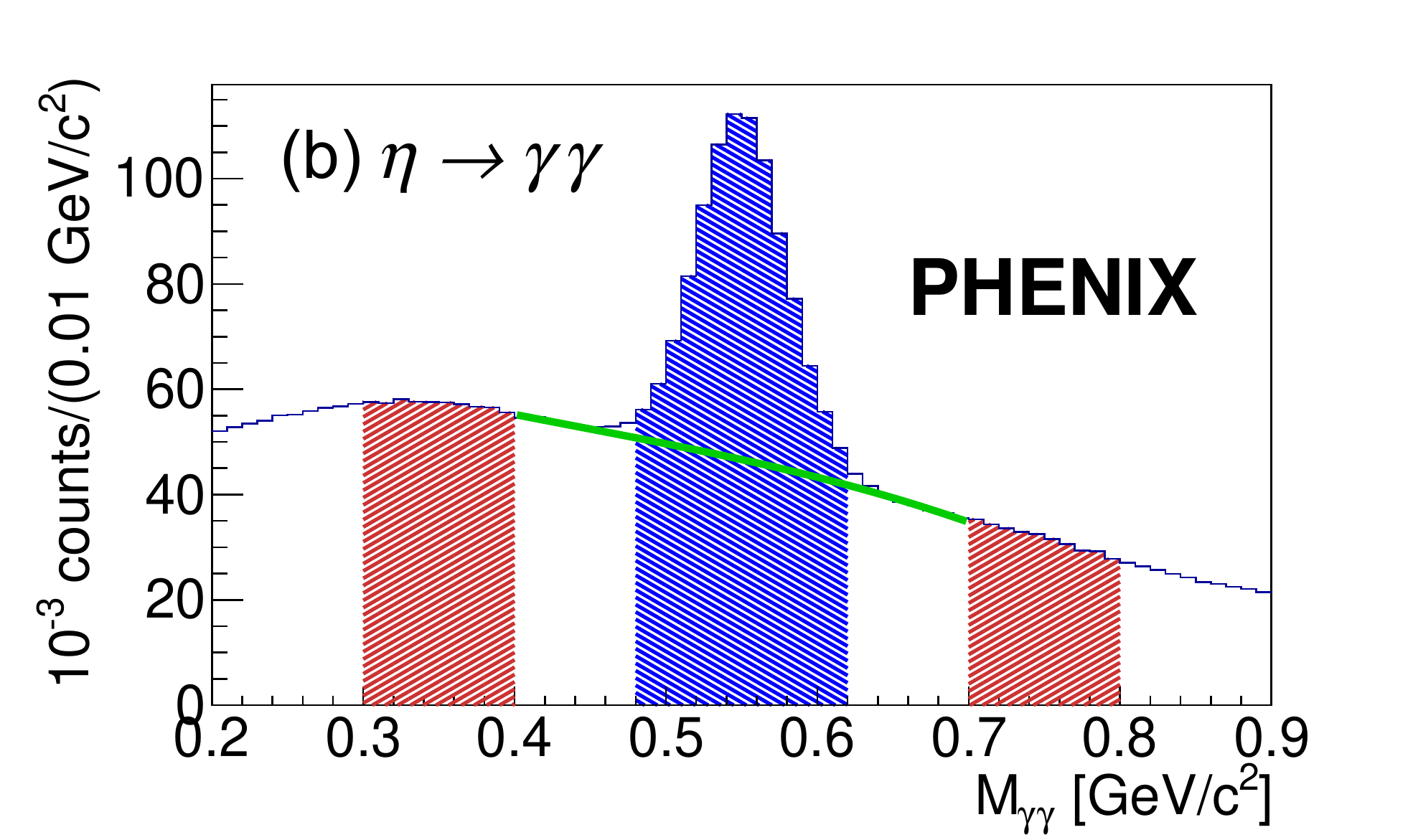}
\caption{Example invariant mass distributions around the (a) \piz  
and (b) $\eta$ peak for photon pairs with ${4<p_T<5 \gevc}$ in one of 
the detector arms.  The minus-$45^{\circ}$ hatched [blue] region at the 
center of each plot corresponds to the invariant mass region under 
the peak which is used to calculate $A_N$ in \eq{eq:sb} and the 
plus-$45^{\circ}$ [red] side band regions 
correspond to the photon pairs that are used to calculate 
$A_N^{\rm BG}$.  The bold solid [green] curves correspond to the fit 
to the combinatorial background, which is used to calculate 
the background fraction.}
  \label{invmass}
\end{figure}

The measured asymmetries are also corrected for background
\begin{equation}
A_N^{\rm Sig} = \frac{A_N - r \cdot A_N^{\rm BG}}{1-r},
\label{eq:sb}
\end{equation}

\noindent where $r$ is the fractional contribution of photon pairs from 
combinatorial background within the invariant mass peak.  The background 
fraction is calculated from fits to the invariant mass spectra where a 
Gaussian is used to describe the invariant mass peak, and a third order 
polynomial is used to describe the combinatorial background, as shown in 
the green curves in \fig{invmass}. Using this method, the contribution 
of combinatorial background under the \piz peak is determined to vary 
from 10\% in the lowest \pt bin to 6\% in the highest. Under the $\eta$ 
meson invariant mass peak, the background fraction varies from 71\% to 
47\% in the lowest to highest \pt bins. In \eq{eq:sb}, the background 
asymmetry, $A_N^{\rm BG}$, is evaluated with photon pairs in side band 
regions located on either side of the signal peak, as represented in the 
red regions in \fig{invmass}.  For the \piz analysis these side band 
regions are $47<M_{\gamma\gamma}<97\mevcc$ and 
$177<M_{\gamma\gamma}<227\mevcc$, and for the $\eta$ meson analysis 
these regions are $300<M_{\gamma\gamma}<400\mevcc$ and 
$700<M_{\gamma\gamma}<800\mevcc$. These background regions match the 
ranges that were used in previous results~\cite{Adare:2013ekj} and are 
chosen to approximate the behavior of the combinatorial background under 
the invariant mass peak. They are selected to be close to the peak while 
still far enough away to contain negligible contributions from signal 
photon pairs. The background asymmetries are consistent with zero across 
all \pt bins. The background asymmetries between the low-mass and 
high-mass regions are also consistent with zero and with each other.

%--------------------------------------------------- Table I
\begin{table*}[thb]
 \caption{\label{table:pi0TSSA}
The measured $A_N$ of \piz in \pp collisions at \sqs=200 GeV as a 
function of \pt. An additional scale uncertainty of 3.4\% due to the 
polarization uncertainty is not shown. The total $\sigma_{\rm syst}$ in 
the lowest \pt bin includes an additional systematic uncertainty of 
$1.06\times10^{-4}$ from bunch shuffling.}
 \begin{ruledtabular}
 \begin{tabular}{ccccccc}
  $\langle\pt\rangle $ & \pt bin ranges & $A_N$ & $\sigma_{\rm stat}$ 
  & $\sigma_{\rm syst}$ & $\sigma_{\rm syst}$ & $\sigma_{\rm syst}$ \\
   $({\rm GeV}/c)$  & $({\rm GeV}/c)$  &  & & (rel.~lumi.~vs sqrt.) & (bg.~fraction) &(total) \\\hline
2.58  & 2--3 & $ 1.43\times10^{-4}$ & $2.81\times10^{-4}$ & $5.71\times10^{-5}$ & $3.92\times10^{-7}$ & $1.20\times10^{-4}$ \\
3.42  & 3--4 & $-3.43\times10^{-4}$ & $3.21\times10^{-4}$ & $1.73\times10^{-5}$ & $3.92\times10^{-6}$ & $1.77\times10^{-5}$ \\
4.40  & 4--5 & $ 3.35\times10^{-4}$ & $5.71\times10^{-4}$ & $6.56\times10^{-5}$ & $1.91\times10^{-6}$ & $6.57\times10^{-5}$ \\
5.40  & 5--6 & $ 2.33\times10^{-3}$ & $1.06\times10^{-3}$ & $9.61\times10^{-5}$ & $6.68\times10^{-7}$ & $9.61\times10^{-5}$ \\
6.41  & 6--7 & $-6.89\times10^{-4}$ & $1.87\times10^{-3}$ & $1.12\times10^{-4}$ & $2.11\times10^{-5}$ & $1.14\times10^{-4}$ \\
7.42  & 7--8 & $ 1.93\times10^{-3}$ & $3.11\times10^{-3}$ & $3.41\times10^{-4}$ & $7.61\times10^{-5}$ & $3.50\times10^{-4}$ \\
8.43  & 8--9 & $-2.38\times10^{-3}$ & $4.88\times10^{-3}$ & $2.45\times10^{-4}$ & $3.99\times10^{-4}$ & $4.69\times10^{-4}$ \\
9.43  & 9--10 & $ 4.04\times10^{-4}$ & $7.03\times10^{-3}$ & $3.31\times10^{-4}$ & $1.16\times10^{-4}$ & $3.51\times10^{-4}$ \\
10.79 & 10--12 & $ 7.34\times10^{-3}$ & $7.99\times10^{-3}$ & $9.71\times10^{-5}$ & $3.13\times10^{-4}$ & $3.28\times10^{-4}$ \\
13.53 & 10--20 & $-1.05\times10^{-2}$ & $1.27\times10^{-2}$ & $6.86\times10^{-4}$ & $1.15\times10^{-5}$ & $6.86\times10^{-4}$ \\
 \end{tabular}
 \end{ruledtabular}
% \end{table*}
%--------------------------------------------------- Table II
%\begin{table*}[tbh]
 \caption{\label{table:etaTSSA}
    The measured $A_N$ of $\eta$ mesons in \pp collisions at \sqs=200 GeV as a function of \pt.  An additional scale uncertainty of 3.4\% due to the polarization uncertainty is not shown. The total $\sigma_{\rm syst}$ in the lowest \pt bin includes an additional systematic uncertainty of $6.20\times10^{-4}$ from bunch shuffling. }
 \begin{ruledtabular}
 \begin{tabular}{ccccccc}
     $\langle\pt\rangle $ & \pt bin edges & $A_N$ & $\sigma_{\rm stat}$ 
     & $\sigma_{\rm syst}$  & $\sigma_{\rm syst}$  & $\sigma_{\rm syst}$ \\
     $({\rm GeV}/c)$ &  $({\rm GeV}/c)$ & & & (rel.~lumi.~vs sqrt.) & (bg.~fraction) &(total) \\\hline
2.39  & 2--3 & $ 2.44\times10^{-3}$ & $1.83\times10^{-3}$ & $5.18\times10^{-4}$ & $4.58\times10^{-5}$ & $8.09\times10^{-4}$ \\
3.53  & 3--4 & $-1.99\times10^{-3}$ & $1.59\times10^{-3}$ & $8.36\times10^{-5}$ & $3.31\times10^{-5}$ & $8.99\times10^{-5}$ \\
4.39  & 4--5 & $-3.31\times10^{-3}$ & $2.48\times10^{-3}$ & $1.44\times10^{-4}$ & $4.55\times10^{-5}$ & $1.51\times10^{-4}$ \\
5.40  & 5--6 & $-1.39\times10^{-3}$ & $4.21\times10^{-3}$ & $2.41\times10^{-4}$ & $3.59\times10^{-5}$ & $2.44\times10^{-4}$ \\
6.41  & 6--7 & $ 2.22\times10^{-3}$ & $7.09\times10^{-3}$ & $1.12\times10^{-3}$ & $6.35\times10^{-6}$ & $1.12\times10^{-3}$ \\
7.42  & 7--8 & $ 1.03\times10^{-2}$ & $1.15\times10^{-2}$ & $7.03\times10^{-4}$ & $1.60\times10^{-4}$ & $7.20\times10^{-4}$ \\
8.75  & 8--10 & $ 7.90\times10^{-3}$ & $1.37\times10^{-2}$ & $1.24\times10^{-3}$ & $1.88\times10^{-4}$ & $1.25\times10^{-3}$ \\
11.76 & 10--20 & $ 1.68\times10^{-2}$ & $2.19\times10^{-2}$ & $4.25\times10^{-3}$ & $3.70\times10^{-4}$ & $4.26\times10^{-3}$ \\
 \end{tabular}
 \end{ruledtabular}
 \end{table*}

Tables~\ref{table:pi0TSSA}~and~\ref{table:etaTSSA} show the asymmetries 
with statistical and systematic uncertainties.  The total systematic 
uncertainty is the sum of the three sources of systematic uncertainty 
added in quadrature.  The systematic uncertainty on the asymmetry due to 
the background fraction in \eq{eq:sb} is determined by varying the fit 
ranges when computing $r$ and calculating how much the 
background-corrected asymmetry changes.  While the asymmetries 
calculated with the ``relative luminosity" [\eq{eq:relan}] and the 
``square root" [\eq{eq:sqrtan}] formulas were found to be statistically 
consistent, their difference was assigned as a conservative systematic 
uncertainty due to possible variations in detector performance and beam 
conditions. This dominates the total systematic uncertainty for most \pt 
bins.

Bunch shuffling is a technique used to investigate potential sources of 
systematic uncertainty that could cause the measured asymmetry results 
to vary from their true values beyond statistical fluctuations.  Bunch 
shuffling involves randomizing the assigned bunch-by-bunch polarization 
directions of the beam such that the physical asymmetry disappears, 
thereby isolating the statistical variations present in the data.  All 
asymmetry values have bunch shuffling results consistent with 
statistical variations except for the lowest \pt bin where there is 7\% 
and 6\% more variation beyond what is expected from statistical 
fluctuations in the \piz and $\eta$ meson analyses, respectively.  These 
values are used to assign additional systematic uncertainties to the 
lowest \pt bin of the \piz and $\eta$ meson asymmetries and dominate the 
total systematic uncertainty for those bins.

Additional cross checks included examining the asymmetries in the two 
arms separately using \eq{eq:relan} and measuring the asymmetry as an 
explicit function of $\phi$. All checks were statistically consistent 
with the main asymmetry results.

\section{Results and discussion}

\begin{figure}[tbh]
  \includegraphics[width=1.0\linewidth]{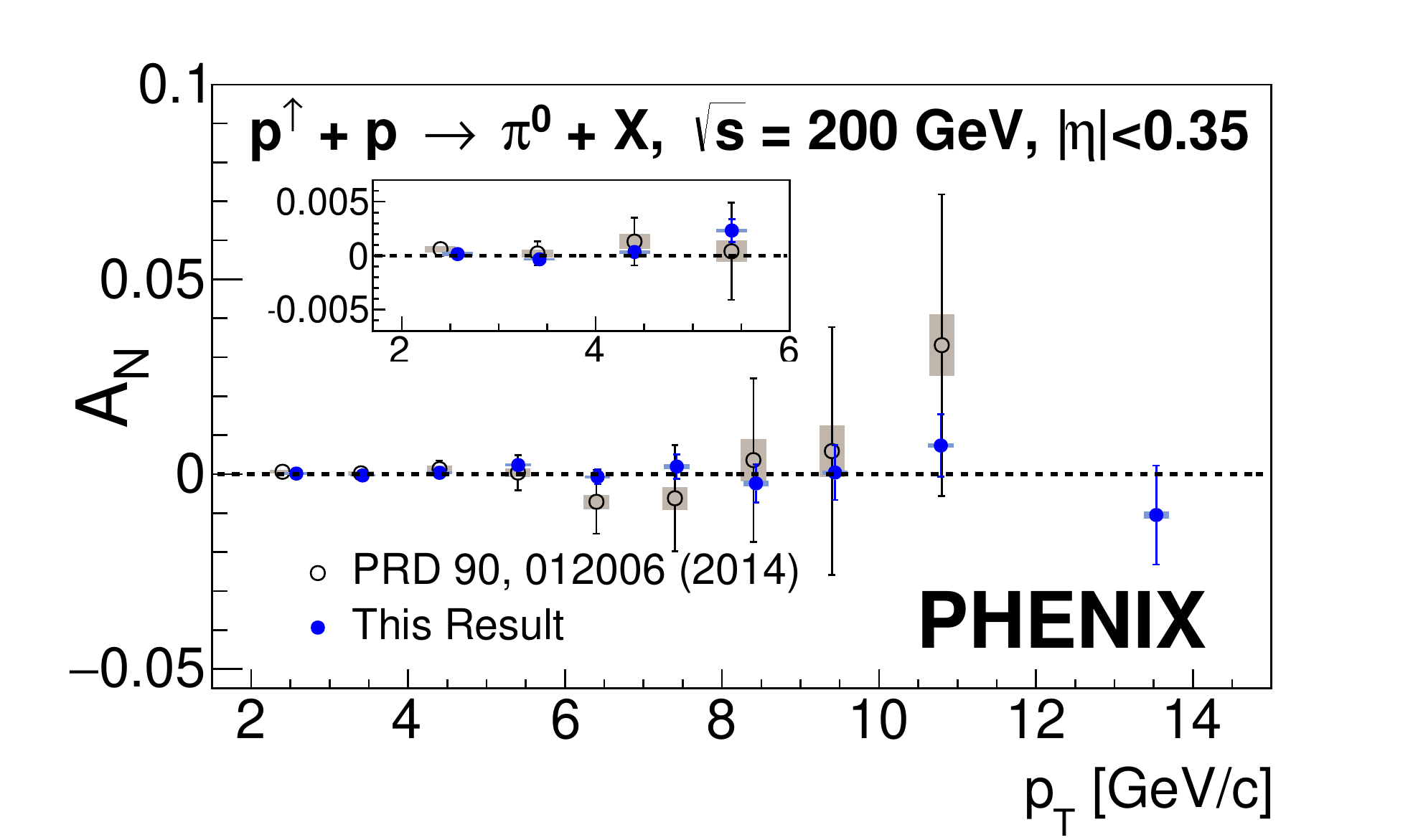}
\caption{Transverse single-spin asymmetry of neutral pions measured at 
$|\eta|<0.35$ in \ptp collisions at $\sqrt{s} = 200 \gev$. An additional 
scale uncertainty of 3.4\% due to the polarization uncertainty is not 
shown.}
  \label{fig:pionAN}
%\end{figure}

%%%%%%%%%%%%%
% Fig. AN vs pT               %
%%%%%%%%%%%%%

%%%%%%%%%%%%%%%%%%%%%%%%%%%%%%%%%%%%%%%%%%%%%%%%%% Fig_3
%\begin{figure}[tbh]
  \includegraphics[width=1.0\linewidth]{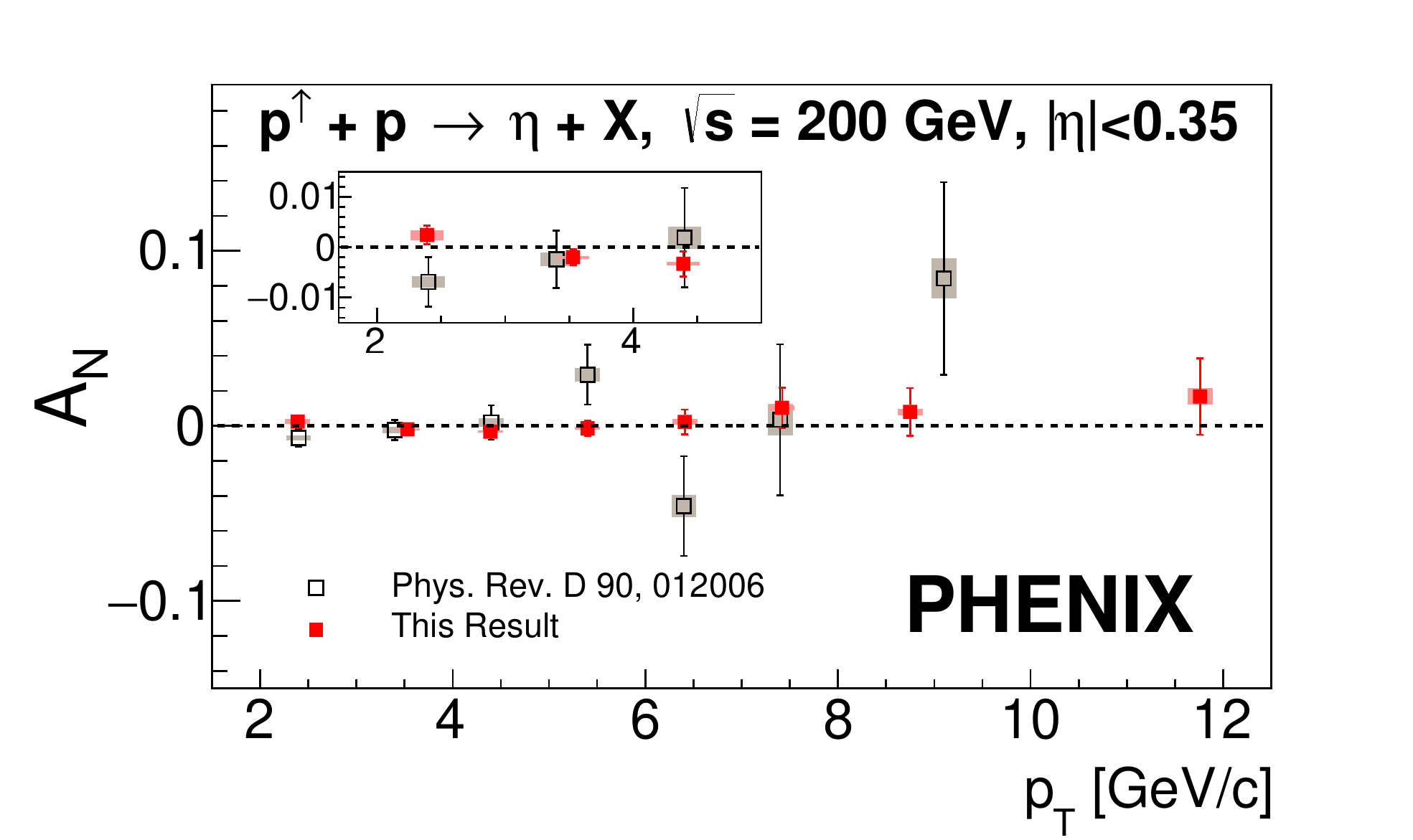}
\caption{Transverse single-spin asymmetry of eta mesons measured at 
$|\eta|<0.35$ in \ptp collisions at $\sqrt{s} = 200 \gev$. An additional 
scale uncertainty of 3.4\% due to the polarization uncertainty is not 
shown.}
  \label{fig:etaAN}
\end{figure}

Figure~\ref{fig:pionAN} shows the $A_N$ of neutral pions at midrapidity 
in \ptp collisions at \sqs=200~GeV, where the bands represent the 
systematic uncertainty and the bars represent the statistical 
uncertainty. The comparison to previous results~\cite{Adare:2013ekj} 
demonstrates the improvement in statistical precision.  The inset in 
\fig{fig:pionAN} shows a zoomed-in comparison at small \pt. The new 
measurement is consistent with our previous measurement and improves the 
precision on average by a factor of 3. The new measurement of $A_N$ of 
neutral pions is consistent with zero in the entire \pt range.

%%%%%%%%%%%%%
% Fig. AN vs pT               %
%%%%%%%%%%%%%

%%%%%%%%%%%%%%%%%%%%%%%%%%%%%%%%%%%%%%%%%%%%%%%%%% Fig_2

\begin{figure}[tbh]
  \includegraphics[width=1.0\linewidth]{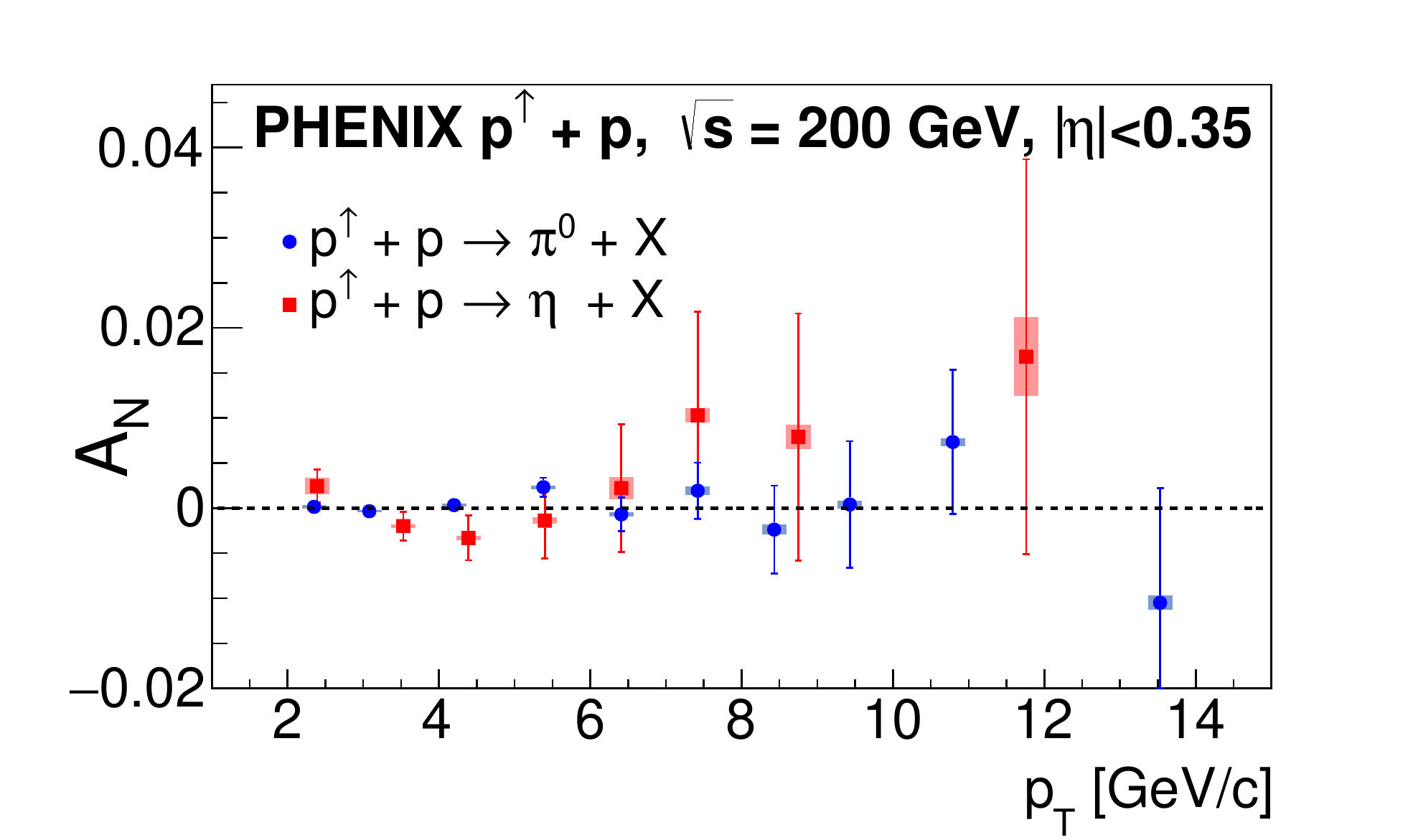}
\caption{Comparison of the \piz and $\eta$ meson asymmetries 
measured at $|\eta|<0.35$ in \ptp collisions at $\sqrt{s}=200\gev$. 
An additional scale uncertainty of 3.4\% due to the polarization 
uncertainty is not shown.}
  \label{fig:pi0etaAN}
\end{figure}

The measurement of $A_N$ of $\eta$ mesons in \ptp collisions at \sqs = 
200 GeV is shown in \fig{fig:etaAN}. This measurement is also compared 
to the previous result, similarly to \fig{fig:pionAN}. The new 
measurement is consistent with the previous result and with zero across 
the entire \pt range. In principle, comparisons of \piz and $\eta$ meson 
TSSAs may indicate additional effects from strange quarks, isospin 
differences, or hadron mass. At forward rapidity, existing 
measurements~\cite{Adare:2014qzo, Adamczyk:2012xd} do not yet clearly 
resolve whether the $\eta$ meson asymmetry is larger than the \piz 
asymmetry as predicted in some models~\cite{Kanazawa:2011bg}. At 
midrapidity, there is a larger contribution from gluon dynamics and, as 
shown in \fig{fig:pi0etaAN}, both asymmetries are consistent with zero 
and therefore show no evidence for differences due to strangeness, 
isospin, or mass.

%%%%%%%%%%%%%
% Fig. AN vs pT               %
%%%%%%%%%%%%%

%%%%%%%%%%%%%%%%%%%%%%%%%%%%%%%%%%%%%%%%%%%%%%%%%% Fig_4

%%%%%%%%%%%%%
% Fig. AN vs pT               %
%%%%%%%%%%%%%

%%%%%%%%%%%%%%%%%%%%%%%%%%%%%%%%%%%%%%%%%%%%%%%%%% Fig_5
\begin{figure}[tbh]
  \includegraphics[width=1.0\linewidth]{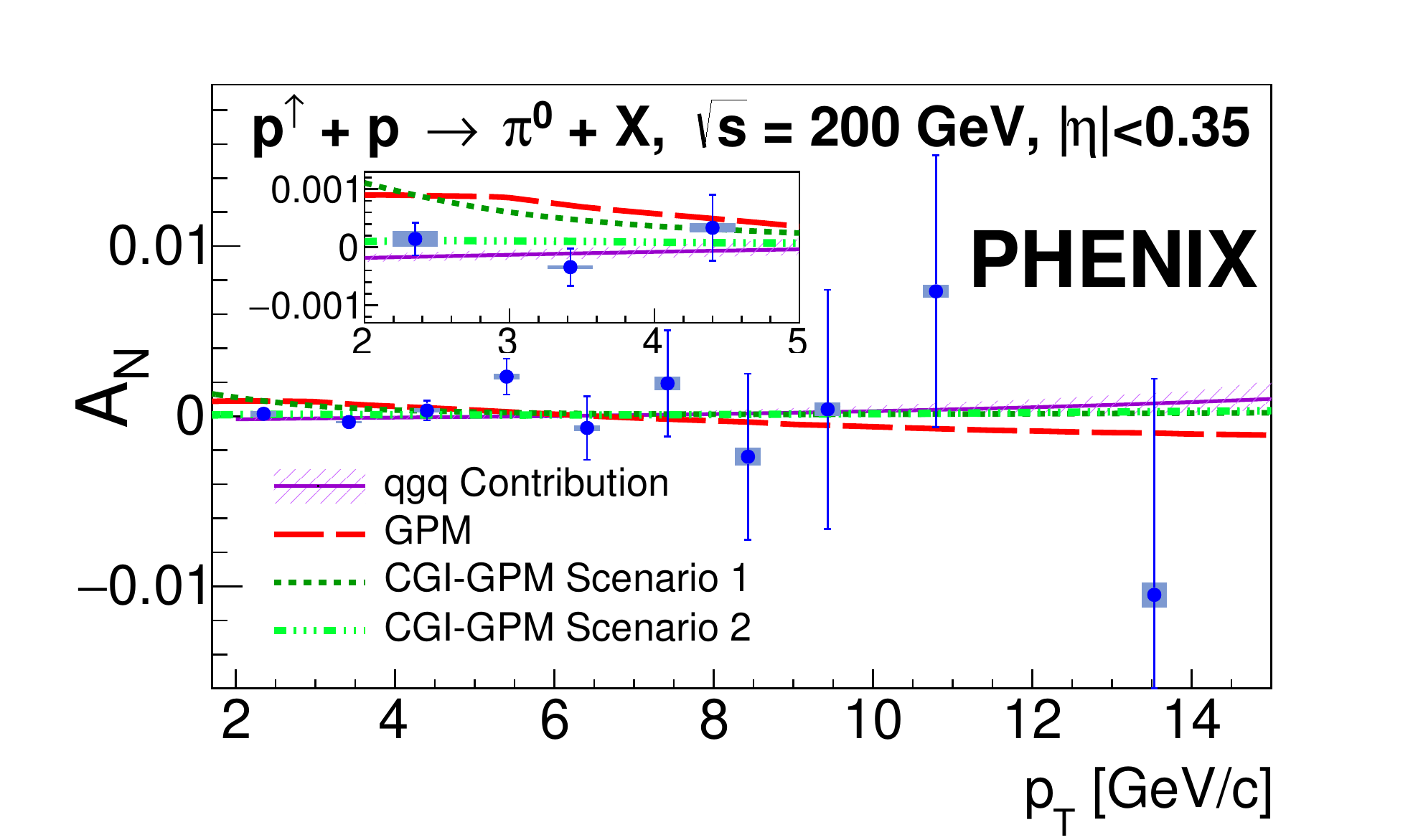}
\caption{This \piz asymmetry result plotted with theory calculations for 
the asymmetry in both the collinear twist-3~\cite{Cammarota:2020qcw} and 
TMD~\cite{DAlesio:2018rnv} theoretical frameworks. See text for details. 
}
  \label{fig:pionANtheory}
\end{figure}

Figure~\ref{fig:pionANtheory} shows this \piz TSSA result plotted with 
theoretical predictions.  The \textit{qgq} curve shows the predicted 
contribution from collinear twist-3 \textit{qgq} functions from both the 
polarized proton and the process of hadronization.  This curve was 
calculated with fits that were published in 
Ref.~\cite{Cammarota:2020qcw} and has been reevaluated in the rapidity 
range of PHENIX.  Midrapidity \piz production includes a large 
fractional contribution from gluons in the proton, so a complete 
collinear twist-3 description of the midrapidity \piz TSSA would also 
need to include the contribution from the trigluon correlation function. 
Given the small expected contribution from the \textit{qgq} correlation 
function, this measurement can constrain future calculations of the 
\textit{ggg} correlation function, such as those in 
Ref.~\cite{Beppu:2013uda}.

The other theory curves in \fig{fig:pionANtheory} show predictions for 
the midrapidity \piz TSSA generated by the Sivers TMD PDF.  These curves 
include contributions from both the quark and gluon Sivers functions and 
have been evaluated for $x_F = 0$, which approximates the measured 
kinematics. These calculations use the generalized parton model (GPM) 
which takes the first $k_T$ moment of the Sivers function (e.g. 
$\int{k_T\cdot q(k_T)}$) and does not include next-to-leading-order 
interactions with the proton fragments.  The ``GPM'' curve uses the 
parameters stated in Eq.~(32) of Ref.~\cite{DAlesio:2018rnv}.  The 
color-gauge-invariant generalized parton model (CGI-GPM) expands on the 
GPM by including initial- and final-state interactions through the 
one-gluon exchange approximation.  This model has been shown to 
reproduce the predicted sign change for the quark Sivers function in 
SIDIS and Drell-Yan. The CGI-GPM curves plotted 
in~\fig{fig:pionANtheory} show two different scenarios for this model, 
the specifics of which can be found in Eq.~(34) of 
Ref.~\cite{DAlesio:2018rnv}. The values that are used for the Scenario 1 
curve are chosen to maximize the open heavy flavor TSSA generated by the 
gluon Sivers function while still keeping this asymmetry within the 
statistical error bars of the published result in 
Ref.~\cite{Aidala:2017pum} and simultaneously describing the previously 
published midrapidity \piz TSSA from Ref.~\cite{Adare:2013ekj}.  The 
values used in the Scenario 2 curve are similarly calculated, except 
that they minimize the open heavy flavor TSSA within the range of the 
published statistical error bars. As shown in the zoomed-in inset of 
\fig{fig:pionANtheory}, this \piz TSSA result has the statistical 
precision at low \pt needed to distinguish between the GPM and CGI-GPM 
frameworks, preferring CGI-GPM Scenario 2.

Measurements of TSSAs in \pp collisions are essential to understanding 
the underlying nonperturbative processes which generate them.  In 
particular, further measurements are necessary to clarify certain 
questions in the interpretations of the TSSAs. For example, the small 
forward jet asymmetries measured in Ref.~\cite{Bland:2013pkt} have been 
interpreted as a cancellation of up and down quark asymmetries, implying 
that the comparatively forward large neutral pion asymmetries include 
significant contributions from spin-momentum correlations in 
hadronization~\cite{Kanazawa:2014dca}.  Additionally, the $p_T$ 
dependence of these forward rapidity measurements remains to be clearly 
understood; measurements of nonzero asymmetries out to even higher $p_T$ 
would help confirm that these twist-3 observables eventually fall off 
with increasing hard scale.  While the midrapidity measurements here are 
all consistent with zero, they still provide the highest available 
statistical precision and $p_T$ reach available at the PHENIX 
experiment.  While forward rapidity light hadron TSSAs are dominated by 
valence quark spin-momentum correlations in the polarized proton, these 
midrapidity TSSA measurements are sensitive to both quark and gluon 
dynamics at leading order.  Thus these data also provide further 
constraints to gluon spin-momentum correlations in transversely 
polarized protons~\cite{Beppu:2013uda, DAlesio:2015fwo}. 

\section{Summary}

The measurements presented here were motivated by the outstanding 
questions regarding the physical origin of TSSAs. The TSSAs of \piz and 
$\eta$ mesons were measured at midrapidity in \pp collisions at 
$\sqs=200$~\gev by the PHENIX experiment.  The measured \piz ($\eta$) 
meson asymmetry is consistent with zero in the presented \pt range, up 
to precision of $3\times10^{-4}$ $(2\times10^{-3})$ in the lowest \pt 
bins. Both measurements have a significant reduction in uncertainty from 
previous measurements at midrapidity at RHIC.  These data extend 
previous constraints to any presence of gluon spin-momentum correlations 
in transversely polarized protons.

%%%%%%%%%%%%%%%%%%%%%%  ACKNOWLEDGMENTS}  %%%%% MGS19 version
%% 2018 change in Korea

\begin{acknowledgments}

We thank the staff of the Collider-Accelerator and Physics
Departments at Brookhaven National Laboratory and the staff of
the other PHENIX participating institutions for their vital
contributions.  
We also thank D. Pitonyak, S. Yoshida, U. D'Alesio, F. Murgia 
and C. Pisano for helpful discussions.
We acknowledge support from the
Office of Nuclear Physics in the
Office of Science of the Department of Energy,
the National Science Foundation,
Abilene Christian University Research Council,
Research Foundation of SUNY, and
Dean of the College of Arts and Sciences, Vanderbilt University
(U.S.A),
Ministry of Education, Culture, Sports, Science, and Technology
and the Japan Society for the Promotion of Science (Japan),
Conselho Nacional de Desenvolvimento Cient\'{\i}fico e
Tecnol{\'o}gico and Funda\c c{\~a}o de Amparo {\`a} Pesquisa do
Estado de S{\~a}o Paulo (Brazil),
Natural Science Foundation of China (People's Republic of China),
Croatian Science Foundation and
Ministry of Science and Education (Croatia),
Ministry of Education, Youth and Sports (Czech Republic),
Centre National de la Recherche Scientifique, Commissariat
{\`a} l'{\'E}nergie Atomique, and Institut National de Physique
Nucl{\'e}aire et de Physique des Particules (France),
Bundesministerium f\"ur Bildung und Forschung, Deutscher Akademischer
Austausch Dienst, and Alexander von Humboldt Stiftung (Germany),
J. Bolyai Research Scholarship, EFOP, the New National Excellence
Program ({\'U}NKP), NKFIH, and OTKA (Hungary),
Department of Atomic Energy and Department of Science and Technology
(India),
Israel Science Foundation (Israel),
Basic Science Research and SRC(CENuM) Programs through NRF
funded by the Ministry of Education and the Ministry of
Science and ICT (Korea).
Physics Department, Lahore University of Management Sciences (Pakistan),
Ministry of Education and Science, Russian Academy of Sciences,
Federal Agency of Atomic Energy (Russia),
VR and Wallenberg Foundation (Sweden),
the U.S. Civilian Research and Development Foundation for the
Independent States of the Former Soviet Union,
the Hungarian American Enterprise Scholarship Fund,
the US-Hungarian Fulbright Foundation,
and the US-Israel Binational Science Foundation.

\end{acknowledgments}

%%%%%%%%%%%%%%%%%%%%%%%%%%%  References 

%\bibliography{ppg234x1}   

 
\end{document}